\documentclass[aps,PRApplied,twocolumn,superscriptaddress,floatfix]{revtex4}
\usepackage{graphicx}
\usepackage{latexsym}
\usepackage{stmaryrd}
\usepackage{amsmath}
\usepackage{amssymb}
\usepackage{float}
\usepackage[colorlinks,linkcolor=blue,anchorcolor=blue,urlcolor=blue,citecolor=blue]{hyperref}
\begin{document}
\title{Electrical Detection of Ferroelectric-like Metals through Nonlinear Hall Effect}

\author{Rui-Chun Xiao}
\affiliation{School of Physical Science and Technology, Soochow University, Suzhou, 215006, China}
\affiliation{Institute for Advanced Study, Soochow University, Suzhou 215006, China}

\author{Ding-Fu Shao} \email{dfshao@unl.edu}
\affiliation{Department of Physics and Astronomy \& Nebraska Center for Materials and Nanoscience, University of Nebraska, Lincoln, Nebraska 68588-0299, USA}

\author{ Wenjuan Huang}
\affiliation{Changzhou Institute of Technology, Changzhou 213031, China}

\author{Hua Jiang}
\email{jianghuaphy@suda.edu.cn}
\affiliation{School of Physical Science and Technology, Soochow University, Suzhou, 215006, China}
\affiliation{Institute for Advanced Study, Soochow University, Suzhou 215006, China}

\begin{abstract}
Ferroelectric-like metals are a relatively rare class of materials that have ferroelectric-like distortion and metallic conductivity. LiOsO$_3$ is the first demonstrated and the most investigated ferroelectric-like metal. The presence of free carriers makes them difficult to be studied by traditional ferroelectric techniques. In this paper, using the symmetry analysis and first-principles calculations, we demonstrate that the ferroelectric-like transition of LiOsO$_3$ can be probed by a kind of electrical transport method based on nonlinear Hall effect. The Berry curvature dipole exists in the ferroelectric-like phase, and it can lead to a measurable nonlinear Hall conductance with a conventional experimental setup. However, the symmetry of the paraelectric-like phase LiOsO$_3$ vanishes the Berry curvature dipole. The Berry curvature dipole shows a strong dependence on the polar displacement, which might be helpful for the detection of polar order. The nonlinear Hall effect provides an effective method for the detection of phase transition in the study of the ferroelectric-like metals and promotes them to be applied in the ferroelectric-like electronic devices.

\end{abstract}
\maketitle
\section{Introduction}
Ferroelectrics are a kind of crystalline materials that exhibit electrically switchable electrical polarization. In these types of materials, the structural phase transitions happen at the ferroelectric critical temperature, bringing the spontaneous polar atomic distortion and ferroelectric polarization at low temperatures. The ferroelectric materials are usually insulators. In the 1960s, Anderson and Blount pointed out that in a metallic system, a ferroelectric-like structural phase transition can also emerge to introduce the long-range polar order \cite{RN2383}. Due to the coexistence of two seemingly incompatible properties: ferroelectric-like distortion and metallic conductivity, ferroelectric-like metals are a relatively rare class of materials. This concept was first found in LiOsO$_3$ \cite{RN2330}, and then was found in NdNiO$_3$ thin-film \cite{RN2400} and some two-dimensional Van der Waals materials such as WTe$_2$ \cite{RN2306, RN2355} and MoTe$_2$ \cite{RN2525}, \emph{etc.}. The combination of metallicity and polar structures gives rise to a series of unique physical properties, such as unconventional Cooper pairing \cite{RN2414, RN2410, RN2352, RN2413}, highly anisotropic thermopower response \cite{RN2413}, anomalous optical properties \cite{RN2415, RN2411}, and magnetoelectricity \cite{RN2414, RN2492, RN2491}.

Despite these remarkable signs of progress, the efficient detection of the polar order in ferroelectric-like metal remains a challenging problem. The screen of conduction electrons makes it difficult to be characterized by traditional techniques like piezo-force microscopy and ferroelectric hysteresis loop measurements. Structural characterization methods such as X-ray and neutron diffractions can be used to detect the ferroelectric-like phase transition \cite{RN2330}. However, it is difficult to distinguish whether the polar displacement is positive or negative with these methods. The optical techniques such as nonlinear optical method (second harmonic generation) \cite{RN2400, RN2405, RN2333} can probe the structural asymmetry in ferroelectric-like metals. However, the strong absorption of light by metals limits the detection within small thickness below the surface. An efficient electrical transport method to detect the ferroelectric-like transition and polar order would be desirable for the ferroelectric-like metals.

This difficulty might be solved by the recently discovered nonlinear Hall effect \cite{RN2111, RN2146}. Unlike the linear anomalous Hall effect that only appears in the magnetic materials where the time-reversal symmetry $\hat{T}$ is broken, the nonlinear Hall effect can emerge in nonmagnetic (\emph{i.e.} $\hat{T}$ is invariant) materials as a second-order response to an electric field. It requires the broken of inversion symmetry $\hat{I}$ and finite Berry curvature dipole to generate a net anomalous velocity in a metallic system under the application of charge current. This effect has been experimentally observed in few-layer WTe$_2$ \cite{RN2110, RN2105} and monolayer strained MoS$_2$ \cite{RefMoS2}. Since all the ferroelectric-like metals have the noncentrosymmetric polar structure, the nonlinear Hall effect is an intrinsic property of these materials. Therefore, the nonlinear Hall effect can be used to detect the polar order in them.

In this work, we use the most investigated ferroelectric-like metal LiOsO$_3$ as a representative material to demonstrate that the polar order in ferroelectric-like metals can be detected by the nonlinear Hall effect. Based on the symmetry analysis and first-principles density functional theory (DFT) calculations, we predict that the paraelectric-like phase of LiOsO$_3$ prohibits the existence of the nonlinear Hall effect, while the ferroelectric-like LiOsO$_3$ supports a sizable Berry curvature dipole and leads to a measurable nonlinear Hall voltage.

The rest of the paper is arranged as follows. In Sec. ~\ref{sec:Symmetry}, the symmetry analysis of nonlinear Hall effect is performed. In Section ~\ref{sec:Calculation}, we introduce the first-principles calculation methods. The calculation results and the corresponding explanation of the calculation results are shown in Sec. ~\ref{sec:Results}. Discussion is listed in Sec. ~\ref{sec:Discussion}, and a brief conclusion is given in Sec. ~\ref{sec:Conclusion}.

\section{Symmetry analysis}
\label{sec:Symmetry}
An electric field $\mathbf{E}=\text{Re}\left\{E{e}^{i\omega t} \right\}$ with amplitude ${E}$ and frequency $\omega$ can introduce the nonlinear Hall current density \cite{RN2111}
\begin{align}
{{J}_{a}}=\text{Re}\left\{J_{a}^{\left( 0 \right)}+J_{a}^{\left( 2 \right)}{{e}^{i2\omega t}} \right\},
\label{Eq_NHE}
\end{align}
where $a,b,c=\{x,y,z\}$, $J_{a}^{\left( 0 \right)}=\chi _{abc}^{\left( 0 \right)}E_{b}E_{c}^{*}$ describes the rectified current and $J_{a}^{\left( 2 \right)}=\chi _{abc}^{\left( 2 \right)}E_{b}E_{c}$ is the second harmonic current. Here the nonlinear Hall coefficients are
\begin{align}
\chi _{abc}^{\left( 0 \right)}=\chi _{abc}^{\left( 2 \right)}=-{{\epsilon }_{adc}}\frac{{{e}^{3}}\tau {{D}_{bd}}}{2{{\hbar }^{2}}\left( 1+i\omega \tau \right)},
\label{Eq_chi}
\end{align}
which includes the relaxation time $\tau$ and the Berry curvature dipole ${{D}_{bd}}$:
\begin{align}
{{D}_{bd}}=\int \frac{{{d}^{3}}\mathbf{k}}{{{\left( 2\pi  \right)}^{3}}}{{\rho }_{bd}}(\mathbf{k})=-\int{\frac{{{d}^{3}}\mathbf{k}}{{{\left( 2\pi  \right)}^{3}}}\underset{n}{\mathop \sum }\,v_b\Omega_{n\mathbf{k}}^{d}\frac{\partial {{f}_{0}}}{\partial {{E}_{n\mathbf{k}}}}}.
\label{Eq_Dbd}
\end{align}
Here ${{\rho }_{bd}}$ is the Berry curvature dipole density, ${{E}_{n\mathbf{k}}}$ is the energy of the $n$-th band at the $\mathbf{k}$ point, ${{f}_{0}}$ means the equilibrium Fermi-Dirac distribution function, and $\Omega_{n\mathbf{k}}^{d}$ denotes the Berry curvature. Clearly, the factor ${\partial {f}_{0}}/{\partial {{E}_{n\mathbf{k}}}}$ in Eq. (\ref{Eq_Dbd}) indicates that the Berry curvature dipole is a Fermi surface property, and the nonlinear Hall effect can only appear in metallic systems. Similar to the electric dipole in insulators, Berry curvature dipole density is the measure of the separation of positive and negative Berry curvature in metals \cite{RN2111, ruichun2019}.

A nonzero value of ${{D}_{bd}}$ is required to introduce a finite nonlinear Hall current. The Fermi velocity ${{v}_{b}}={\partial {{E}_{n\mathbf{k}}}}/{\partial {{k}_{b}}}$ is odd under both the time-reversal symmetry $\hat{T}$ and the inversion symmetry $\hat{I}$, \emph{i.e.} $\hat{T}{{v}_{b}}\left( \mathbf{k} \right)=-{{v}_{b}}\left( -\mathbf{k} \right)$ and $\hat{I}{{v}_{b}}\left( \mathbf{k} \right)=-{{v}_{b}}\left( -\mathbf{k} \right)$. The Berry curvature is odd under $\hat{T}$ and even under $\hat{I}$, \emph{i.e.} $\hat{T}\Omega_{n\mathbf{k}}^{d}\left( \mathbf{k} \right)=-\Omega_{n\mathbf{k}}^{d}\left( -\mathbf{k} \right)$ and $\hat{I}\Omega_{n\mathbf{k}}^{d}\left( \mathbf{k} \right)=\Omega_{n\mathbf{k}}^{d}\left( -\mathbf{k} \right)$. Therefore, a finite ${{D}_{bd}}$ can emerge in a nonmagnetic (\emph{i.e.} $\hat{T}$ is invariant) material within inversion symmetry $\hat{I}$ broken according to Eq. (\ref{Eq_Dbd}). Furthermore, the existence of Berry curvature dipole needs intrinsic metallic materials with specific symmetry. The constraint of crystal symmetry for the Berry curvature dipole tensor is described by \cite{RN2111}
\begin{align}
D=\det (\hat{S})\hat{S}D{{\hat{S}}^{-1}},
\label{D_symm}
\end{align}
where $\hat{S}$ is the symmetric operation matrix in the crystal point group. Furthermore, Berry curvature dipole $D$ is a traceless tensor \cite{RN2186},
\begin{equation}
\begin{split}
Tr[D]&={{D}_{11}}+{{D}_{22}}+{{D}_{33}}=0.
\label{Tr_Dbd}
\end{split}
\end{equation}
It is because that Berry curvature is a curl quality [${{\Omega }_{n}}(\mathbf{k})={{\nabla }_{\mathbf{k}}}\times {{A}_{n}}\left( \mathbf{k} \right)$, where $A$ is Berry connection].

All the polar point groups (${{C}_{n}}$, $C_{nv}$ $n$=1, 2, 3, 4, 6) can have non-zero Berry curvature dipole, according to Eqs. (\ref{D_symm}) and (\ref{Tr_Dbd}). In these polar groups, it is convenient to define a vector $\mathbf{d}$ \cite{RN2111} as ${{d}_{a}}={{\epsilon }_{abc}}D_{bc}^{-}/2$, where ${{D}^{-}}=\left( D-{{D}^{T}} \right)/2$ is the antisymmetric parts of the Berry curvature dipole tensor. The polar symmetry enforces the finite vector $\mathbf{d}$ oriented along the polar axis \cite{RN2111}, which is something similar to the electric dipole $\mathbf{P}$ in ferroelectric insulators. Ferroelectric-like metals are metallic systems with noncentrosymmetric polar structures. Therefore, the polar phase can be directly reflected by the nonlinear Hall response related to the vector $\mathbf{d}$. Above the critical temperature, these materials transform from the ferroelectric-like phase to the paraelectric-like phase. The paraelectric-like phase has the zero Berry curvature dipole due to the existence of inversion symmetry. This phase transition can change the magnitude of the nonlinear Hall voltage from a finite value to zero. If the polar displacement is switched, the nonlinear Hall voltage will be reversed, because this switching is equivalent to an inversion symmetry operation, which changes the sign of the Berry curvature dipole. This fully shows that the nonlinear Hall effect measurement can be a promising method for the efficient detection of the polar order in the ferroelectric-like metals.

\begin{figure}
\includegraphics[width=0.95\columnwidth]{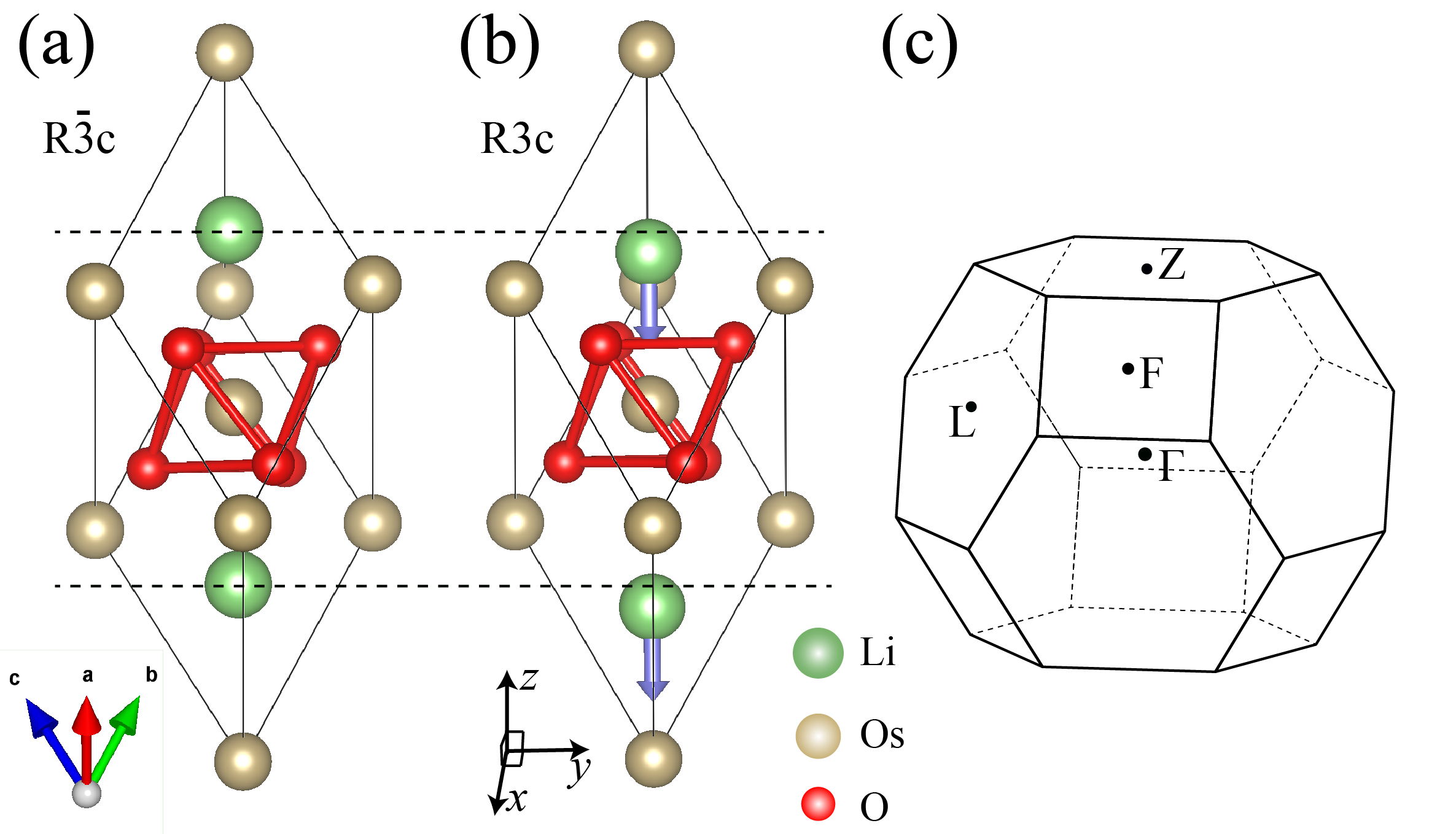}
\caption{Primitive unit cell of (a) paraelectric-like and (b) ferroelectric-like phase of LiOsO$_3$. The ferroelectric-like phase transition mainly involves the displacements of Li atoms. The arrows in (b) stand for the displacements of Li ions along the polar axis ($z$ direction). (c) Brillouin zone of LiOsO$_3$.}
\label{FigCrystal}
\end{figure}

Here we use LiOsO$_3$, the first discovered \cite{RN2330} and most investigated \cite{RN2335, RN2407, RN2297, RN2408, RN2405} ferroelectric-like metal, as a representative example to demonstrate the polar order in ferroelectric-like metals can be detected by the nonlinear Hall effect. At high temperature, LiOsO$_3$ has a centrosymmetric rhombohedral structure with the space group $R\overline{3}c$ (No. 167) and the point group ${{D}_{3d}}$. The Os atom is octahedrally coordinated by six O atoms and located at the center between two Li atoms [Fig. \ref{FigCrystal} (a)]. In this paraelectric-like phase, the Berry curvature dipole is vanishing due to the inversion symmetry $\hat{I}$. A ferroelectric-like structural transition shows up below 140 K \cite{RN2330} due to the ${{A}_{2u}}$ phonon module of the paraelectric-like phase \cite{RN2335, RN2407}, accompanied by mainly the displacement of Li atoms along the polar direction, \emph{i.e.} $z$ ([111]) direction. This displacement results in a ferroelectric-like phase with the space group $R3c$ (No. 161) and the point group ${{C}_{3v}}$, as shown in Fig. \ref{FigCrystal} (b). This point group contains a threefold rotation symmetry ${{C}_{3z}}$ around $z$ direction, and three mirror planes parallel to the $z$ direction. According to Eqs. (\ref{D_symm}) and (\ref{Tr_Dbd}), the Berry curvature dipole tensor under the point group $C_{3v}$ is (see Supplemental Material \cite{RefSM})
\begin{align}
D=\left[ \begin{matrix}
   0 & {{D}_{xy}} & 0  \\
   -{{D}_{xy}} & 0 & 0  \\
   0 & 0 & 0  \\
\end{matrix} \right].
\label{D_C3v}
\end{align}
There are only two independent non-zero antisymmetric elements. This leads to the vector $\mathbf{d}=\left( 0,0,{{d}_{z}}\right)$ oriented along the polar axis with ${{d}_{z}}={\left( {{D}_{xy}}-{{D}_{yx}} \right)}/{2}={{D}_{xy}}$, which is related to a nonlinear Hall current along $z$ direction induced by an in-plane injecting current (see details in Section ~\ref{sec:Results}).

\section{Calculation Details}
\label{sec:Calculation}
First-principles calculations based on DFT are performed with the projector augmented-wave (PAW) method implemented in the VASP \cite{RN1434,RN1433} package. General gradient approximation (GGA) based on the Perdew-Burke-Ernzerhof (PBE) functional is used. The Brillouin zone is sampled with a 16$\times $16$\times $16 mesh of $k$-points. The lattice parameters and atomic positions are fully relaxed until the force on each atom is less than ${{10}^{-4}}$ eV/\text{\AA}. The DFT Bloch wave functions are projected to maximally localized Wannier functions by the Wannier90 code \cite{RN149, RN772}. The Berry curvature and Berry curvature dipole are calculated by the Wannier function interpolation implement in WannierTools software package \cite{RN1186}. In the Berry curvature dipole calculations, the adaptive broadening scheme for $k$-space integration \cite{RN707} is employed. The convergence test is taken, and the $k$ mesh grids of 300$\times$300$\times$300 are adopted.

\section{Results and Explanations}
\label{sec:Results}
The calculated crystal structural parameters and total energies of the LiOsO$_3$ are shown in Table \ref{Tab1}. We find the calculated lattice constants and the atomic displacement are close to the experimental and previously calculated values \cite{RN2297, RN2330, RN2335}. The ferroelectric-like phase of LiOsO$_3$ has the lower energy by 56 meV/cell compared to the paraelectric-like phase. This result is consistent with the experimental observations that the ferroelectric-like phase is the ground state of the LiOsO$_3$ at low temperatures.

\begin{table}
\centering
\caption{Calculated lattice parameters and the polar displacement of the Li atom of the LiOsO$_3$ with the hexagonal representation. The numbers inside the parentheses are the experimental values reported in Ref. \cite{RN2330}.}
\label{Tab1}
\begin{tabular} {p{2.3cm}|p{1.2cm}p{1.2cm}p{1.5cm}p{1.5cm}}
   \hline
   \hline
   &  $a$ (\AA)  &  $c$ (\AA) &  $\Delta {{z}_{Li}}$(\AA) & $\Delta {E}$(meV) \\
   \hline
Paraelectric-like LiOsO$_3$  & 5.164 (5.064) & 13.178 (13.211) &	0 & 0 \\
Ferroelectric-like LiOsO$_3$  & 5.091 (5.046) & 13.362 (13.239) & 0.479 (0.467) & -56 \\
\hline
\hline
\end{tabular}
\end{table}

\begin{figure}[!htpb]
\includegraphics[width=1.00\columnwidth]{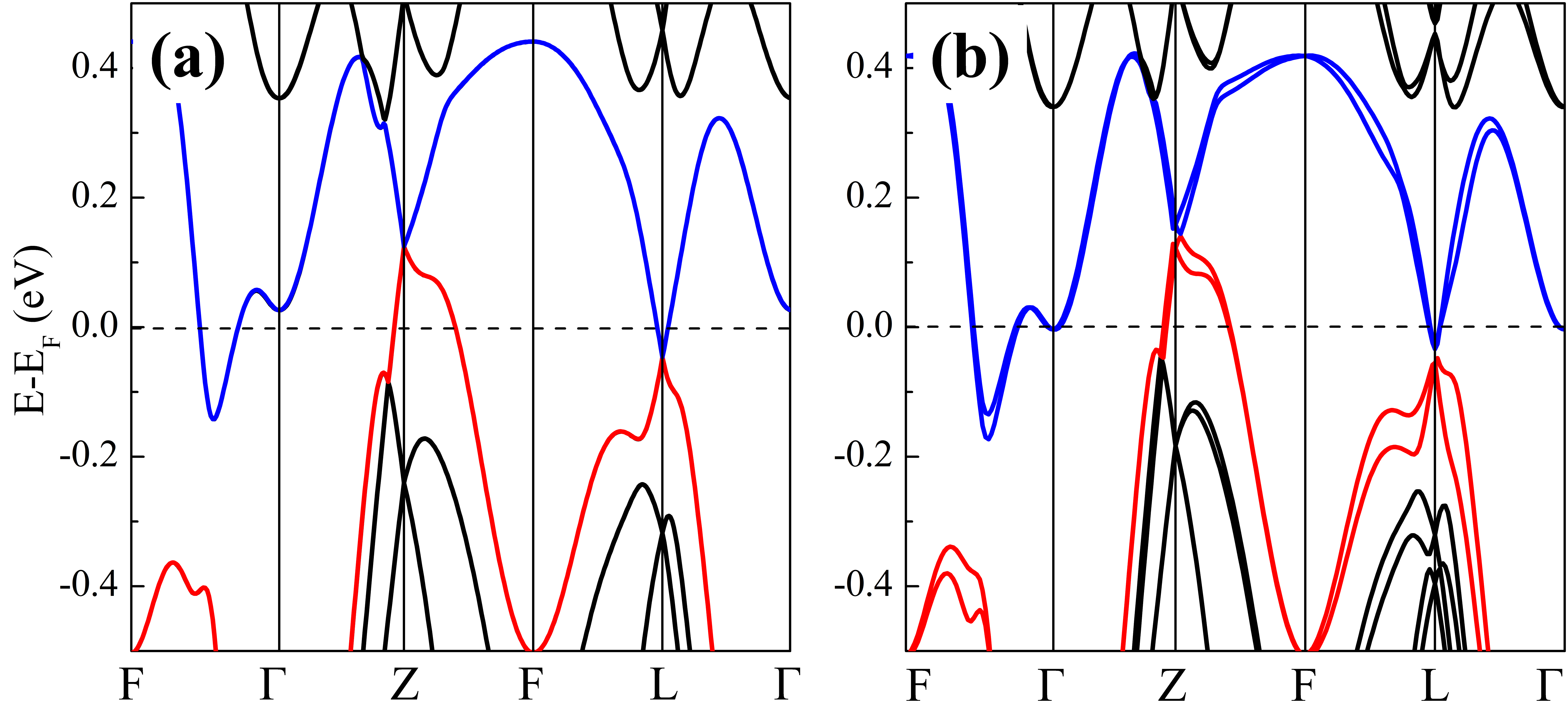}
\caption{Band structure of (a) paraelectric-like and (b) ferroelectric-like LiOsO$_3$.}
\label{FigBand}
\end{figure}

\begin{figure}
\includegraphics[width=1.00\columnwidth]{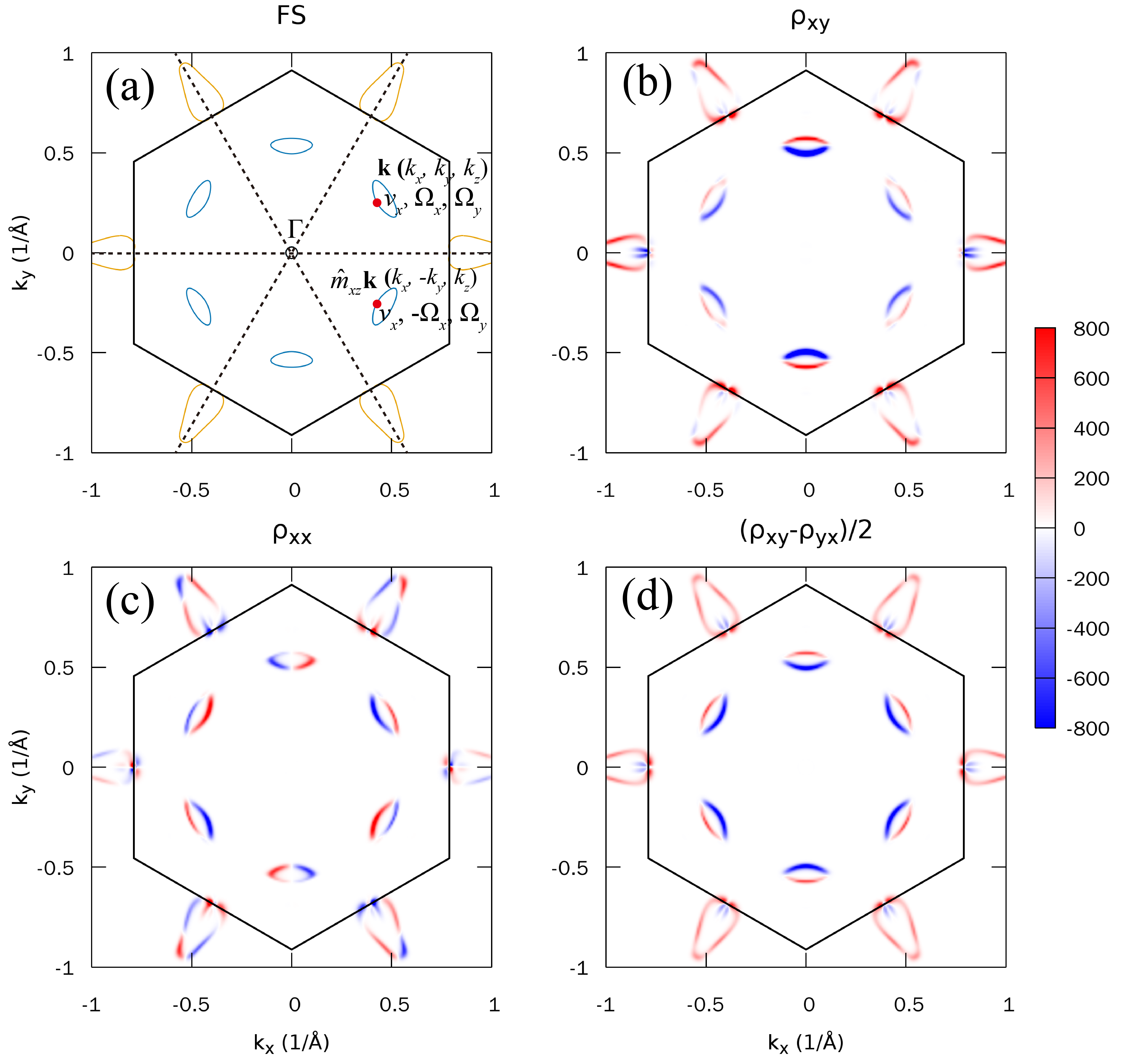}
\caption{(a) Fermi surface of ferroelectric-like LiOsO$_3$. Berry curvature dipole density (in \text{\AA}$^{3}$) of (b) ${{\rho }_{xy}}(\mathbf{k})$, (c) ${{\rho }_{xx}}(\mathbf{k})$, and (d) $\left[ {{\rho }_{xy}}\left( \mathbf{k} \right)-{{\rho }_{yx}}\left( \mathbf{k} \right) \right]/2$ of ferroelectric-like LiOsO$_3$ for ${{k}_{z}}=0$ plane. The dashed lines in (a) denote the mirror planes, and the hexagons mean the Brillouin zone boundary.}
\label{Fig2DBCD}
\end{figure}

The calculated band structures of paraelectric-like and ferroelectric-like LiOsO$_3$ are shown in Figs. \ref{FigBand} (a) and (b), respectively. Consistent with previous theoretical works \cite{RN2297, RN2335}, we find the bands around the Fermi energy ($E_F$) are mainly contributed by the Os-$d$ and O-$p$ orbitals (see Fig. S1 in the Supplemental Material \cite{RefSM}). Since Li atoms are highly ionic and do not bond with Os and O atoms, the polar displacements of Li do not influence the major shape of the band structures around ${{E}_{F}}$. The major difference in the band structures of the two phases is the degeneracy. In the paraelectric-like phase, the presences of the inversion symmetry $\hat{I}$ and time-reversal symmetry $\hat{T}$ enforce the double degeneracy of each band in the Brillouin zone. On the other hand, the inversion symmetry $\hat{I}$ is removed in the ferroelectric-like phase, which destroys the band degeneracy except for the time-reversal $\mathbf{k}$ points. Here we focus on the band structure of the ferroelectric-like phase. There are four bands crossing the ${{E}_{F}}$, forming hole-like pockets at the top and bottom surfaces of the Brillouin zone, and electron-like pockets around the Brillouin zone center and the corners of the Brillouin zone, as shown in Fig. S1 \cite{RefSM}. Figure \ref{Fig2DBCD}(a) shows the cross-section of the Fermi pockets when ${{k}_{z}}=0$, where the threefold rotation $\hat{C}_{3z}$ and three mirror $\hat{m}$ symmetries are clearly reflected.

As described by the symmetry analysis in Sec. ~\ref{sec:Symmetry}, the metallic ground state and the polar structure guarantee the existence of the Berry curvature dipole in ferroelectric-like LiOsO$_3$. This can be seen from the transformations of Berry curvature dipole density $\rho _{bd}$ under symmetry operations reflected by our numerical calculation. For example, the mirror symmetry ${{\hat{m}}_{xz}}$ generates the symmetry transformations of $\mathbf{k}$ piont, velocity and Berry curvature:
${{\hat{m}}_{xz}}\left( {{k}_{x}},{{k}_{y}},{{k}_{z}} \right)=\left( {{k}_{x}},-{{k}_{y}},{{k}_{z}} \right)$, ${{\hat{m}}_{xz}}{{v}_{x}}\left( {{k}_{x}},{{k}_{y}},{{k}_{z}} \right)={{v}_{x}}\left( {{k}_{x}},-{{k}_{y}},{{k}_{z}} \right)$, and ${{\hat{m}}_{xz}}\Omega_{n}^{y}\left( {{k}_{x}},{{k}_{y}},{{k}_{z}} \right)=\Omega_{n}^{y}\left( {{k}_{x}},-{{k}_{y}},{{k}_{z}} \right)$, as shown in Fig. \ref{Fig2DBCD} (a). Therefore, ${{\rho }_{xy}}=-\underset{n}{\mathop \sum }\,{{v}_{x}}\Omega_{n\mathbf{k}}^{y}\frac{\partial {{f}_{0}}}{\partial {{E}_{n\mathbf{k}}}}$ is symmetric with ${{\hat{m}}_{xz}}$ operation, \emph{i.e.} ${{\hat{m}}_{xz}}{{\rho }_{xy}}\left( {{k}_{x}},{{k}_{y}},{{k}_{z}} \right)={{\rho }_{xy}}\left( {{k}_{x}},-{{k}_{y}},{{k}_{z}} \right)$ as shown in Fig. \ref{Fig2DBCD} (b). As a result, the Berry curvature dipole component ${{D}_{xy}}=\int{\frac{{{d}^{3}}\mathbf{k}}{{{\left( 2\pi  \right)}^{3}}}{{\rho }_{xy}}}$ is nonvanishing. For another, the ${{\rho }_{xx}}=-\underset{n}{\mathop \sum }\,{{v}_{x}}\Omega_{n\mathbf{k}}^{x}\frac{\partial {{f}_{0}}}{\partial {{E}_{n\mathbf{k}}}}$ is asymmetrically distributed with respect to the mirror plane ${{\hat{m}}_{xz}}$ [Fig. \ref{Fig2DBCD} (c)] due to ${{\hat{m}}_{xz}}\Omega_{n}^{x}\left( {{k}_{x}},{{k}_{y}},{{k}_{z}} \right)=-\Omega_{n}^{x}\left( {{k}_{x}},-{{k}_{y}},{{k}_{z}} \right)$, which leads to ${{\hat{m}}_{xz}}{{\rho }_{xx}}\left( {{k}_{x}},{{k}_{y}},{{k}_{z}} \right)=-{{\rho }_{xx}}\left( {{k}_{x}},-{{k}_{y}},{{k}_{z}} \right)$. Therefore, ${{D}_{xx}}=\int{\frac{{{d}^{3}}\mathbf{k}}{{{\left( 2\pi \right)}^{3}}}{\rho _{xx}}}=0$. The whole symmetries of velocity and Berry Curvature under $C_{3v}$ symmetry are shown in the Table S1 in the Supplemental Material \cite{RefSM}. Similarly, we can check the symmetry distribution of other ${{\rho }_{bd}}$. The combination of antisymmetric parts of Berry curvature dipole density $\left[ {{\rho }_{xy}}\left( \mathbf{k} \right)-{{\rho }_{yx}}\left( \mathbf{k} \right) \right]/2$ has the threefold symmetry and distributes symmetrically with respect to the three mirror planes, as shown in Fig. \ref{Fig2DBCD} (d), due to ${{\rho }_{xy}}\left( \mathbf{k} \right)-{{\rho }_{yx}}\left( \mathbf{k} \right)={{\rho }_{xy}}\left( \hat{C}_{3z}^{+}\mathbf{k} \right)-{{\rho }_{yx}}\left( \hat{C}_{3z}^{+}\mathbf{k} \right)={{\rho }_{xy}}\left( \hat{C}_{3z}^{-}\mathbf{k} \right)-{{\rho }_{yx}}\left( \hat{C}_{3z}^{-}\mathbf{k} \right)$ (see Supplemental Material \cite{RefSM}). These results indicate that our numerical calculations are consistent with the symmetry. Besides, only the $\mathbf{k}$ points located at the Fermi surface have the nonzero ${{\rho }_{bd}}$ as shown in Fig. \ref{Fig2DBCD}, which implies Berry curvature dipole is indeed the property of Fermi surface.

\begin{figure}[!htpb]
\includegraphics[width=0.9\columnwidth]{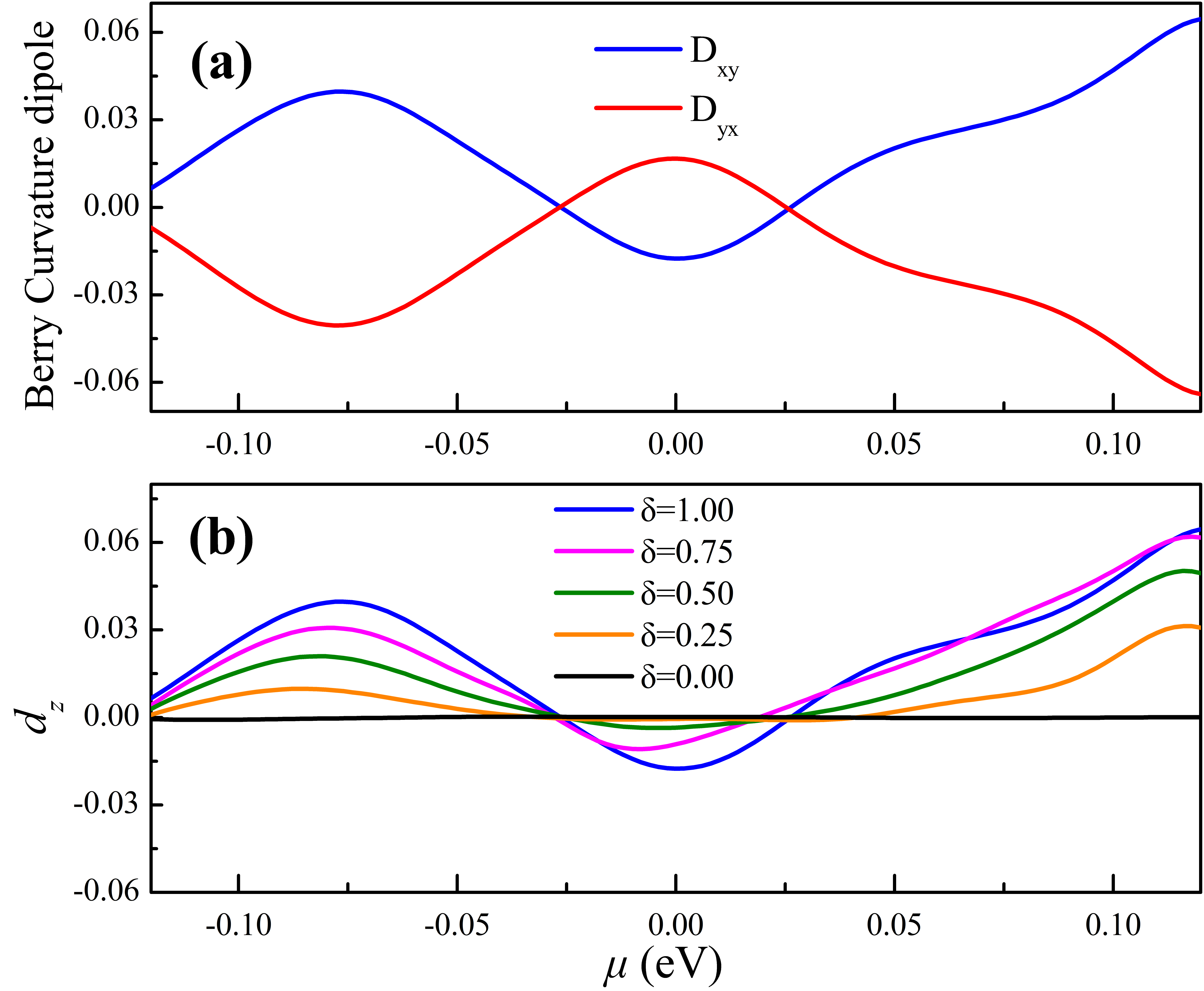}
\caption{(a) ${{D}_{xy}}$ and ${{D}_{yx}}$ as a function of the chemical potential for the ferroelectric-like LiOsO$_3$. (b) The evolution ${{d}_{z}}$ with polar order.}
\label{FigDxy}
\end{figure}

Figure \ref{FigDxy}(a) shows the calculated Berry curvature dipole as a function of the chemical potential. ${{D}_{xy}}=-{{D}_{yz}}$, which is consistent with the above symmetry analysis. The calculated Berry curvature dipole is about 0.017 at ${{E}_{F}}$ and can be enhanced by proper doping. The Berry curvature dipole is comparable to those predicted in Weyl semimetals TaAs, MoTe$_2$ \cite{RN1672}, and topological insulator BiTeI \cite{RN2190}. In order to show the dependence of ${{D}_{xy}}$ on the polar order, we manually move the atoms along the $z$ direction and define $\delta$ as a parameter to reflect the polar displacement. $\delta =0$ corresponds to the paraelectric-like LiOsO$_3$ phase, while $\delta =1$ corresponds to the ferroelectric-like phase. As shown in Fig. \ref{FigDxy} (b), the magnitude of ${{d}_{z}}$ monotonically decreases with $\delta$ and vanishes in the paraelectric-like phase ($\delta =0$). Clearly, the ferroelectric-like phase transition can be reflected by the magnitude change of the nonlinear Hall voltage.

\begin{figure*}[!htp]
\includegraphics[width=0.90\textwidth]{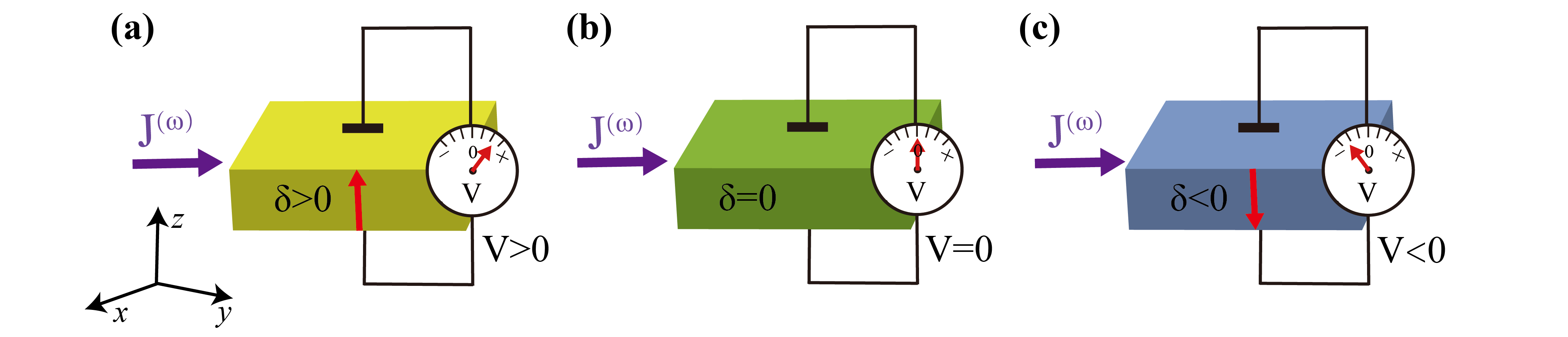}
\caption{(a) Schematics of the polar order detection in ferroelectric-like metal by nonlinear Hall effect. The paraelectric-like and ferroelectric-like state of LiOsO$_3$ are all metallic, so the driving current can go through the $x-y$ plane (perpendicular to the polar axis). An in-plane charge current generates a nonlinear Hall voltage along the out-of-plane direction, which is determined by the polar displacement (denoted by $\delta$). The nonlinear Hall voltage is (a) positive for $\delta >0$, (b) zero for $\delta =0$, and (c) negative for $\delta <0$.}
\label{FigIlldomain}
\end{figure*}

Next, we discuss the nonlinear Hall response induced by the Berry curvature dipole in ferroelectric-like LiOsO$_3$. In the polar group system with Berry curvature dipole $\mathbf{d}$, the nonlinear Hall current density driven by an electrical field $\mathbf{E}$ can be written as \cite{RN2111}
\begin{align}
\left\{ \begin{matrix}
   {{J}^{0}}=\frac{{{e}^{3}}\tau }{2{{\hbar }^{2}}\left( 1+i\omega \tau  \right)}{\mathbf{E}^{*}}\times \left( \mathbf{d}\times \mathbf{E} \right),  \\
   {{J}^{2\omega }}=\frac{{{e}^{3}}\tau }{2{{\hbar }^{2}}\left( 1+i\omega \tau  \right)}\mathbf{E}\times \left( \mathbf{d}\times \mathbf{E} \right).  \\ 
\end{matrix} \right.
\label{Eq_J}
\end{align}
For an electric field $\mathbf{E}=E{{e}^{i\omega t}}\left( \sin \theta \cos \varphi ,\sin \theta \sin \varphi ,\cos \theta  \right)$ ($\theta$ is the polar angle relative to the $z$ axis, and $\varphi $ is the azimuthal angle relative to the $x$ axis), we obtain the induced nonlinear Hall current density
\begin{equation}
\begin{split}
&\left| {{J}^{0}} \right|=\left| {{J}^{2\omega }} \right|=\frac{{{e}^{3}}\tau {{d}_{z}}{{E}^{2}}}{2{{\hbar }^{2}}\left( 1+i\omega \tau  \right)}
\\
&\times \left[ -\cos \theta \sin \theta \cos \varphi ,-\cos \theta \sin \theta \sin \varphi ,{{\sin }^{2}}\theta  \right].
\label{J_dir}
\end{split}
\end{equation}
Clearly, the nonlinear Hall current is absent when the electric field $\mathbf{E}$ is along the $z$ axis ($\theta =0{}^\circ $), according to Eq. (\ref{J_dir}). When the electric field $\mathbf{E}$ is parallel to the $x-y$ plane ($\theta =90{}^\circ$), the in-plane component of the nonlinear Hall current vanishes, while the out-of-plane component exists. Moreover, according to Eq. (\ref{J_dir}) it is independent with the azimuthal angle. This angle-dependence of the nonlinear Hall effect in LiOsO$_3$ is much simpler than that of the nonlinear optics experiment \cite{RN2333}, indicating the convenience to detect the polar order. Besides, different from the nonlinear optical method used in ferroelectric-like metal \cite{RN2400, RN2333}, the nonlinear Hall current flows inside the bulk of metals, which means the influence of the surface is small.

In the DC limit ($\omega \to 0$), the nonlinear Hall conductance induced by an in-plane electric field is
\begin{align}
{{\sigma }^{NHE}}=\left( {{J}^{\left( 0 \right)}}+{{J}^{\left( 2\omega  \right)}} \right)/{E}=\frac{{{e}^{3}}\tau {{d}_{z}}E}{{{\hbar }^{2}}}.
\label{Eq_sigma}
\end{align}
In the constant relaxation-time approximation, the Ohmic conductivity is expressed as \cite{RN707}
\begin{align}
{{\sigma }_{ab}}=\frac{{{e}^{2}}\tau }{{{\hbar}^{2}}}{{C}_{ab}},
\label{Eq_sigma2}
\end{align}
where
\begin{align}
{{C}_{ab}}=\int \frac{1}{{{(2\pi )}^{3}}}\underset{n}{\mathop \sum }\,\frac{\partial {{E}_{n\mathbf{k}}}}{\partial {{k}_{a}}}\frac{\partial {{E}_{n\mathbf{k}}}}{\partial {{k}_{b}}}\left( -\frac{\partial {{f}_{0}}}{\partial {{E}_{n\mathbf{k}}}} \right).
\label{C_ab}
\end{align}
The parameter $C$ can be easily obtained in our DFT calcualtions (see Fig. S2 in the Supplemental Material \cite{RefSM}). Using Eqs. (\ref{Eq_sigma2}) and (\ref{C_ab}), we can drop out of the relaxation-time $\tau$ in Eq. (\ref{Eq_sigma}). Therefore, the nonlinear Hall conductivity in the DC limit can be simply estimated by
\begin{align}
{{\sigma }^{NHE}}=e\frac{{{d}_{z}}}{{{C}_{\bot}}}{{J}}.
\label{Eq_sigma_Cab}
\end{align}
Using the calculated ${{d}_{z}} \sim 0.02$ and ${{C}_{\bot}} \sim$ 0.02 eV/\text{\AA}  (${\bot}$ means the $x-y$ direction, \emph{e.g.} $C_{xx}$) in the ferroelectric-like LiOsO$_3$ and a conventional driving electric current $J \sim 5 \times {{10}^{6}}$ $\text{A}/\text{cm}^{2}$, the calculated nonlinear Hall conductance of ${{\sigma }^{NHE}} \sim$ 5 ${{\Omega}^{-1}}{\text{m}^{-1}}$ is estimated, according to Eq. (\ref{Eq_sigma_Cab}). The sizable nonlinear Hall conductance is comparable to those value of some anomalous Hall materials \cite{RN2353}. To sum up, the nonlinear Hall effect measurement really a promising and efficient detection method for the polar order in LiOsO$_3$.

\section{Discussion}
\label{sec:Discussion}
The manipulation and detection of the magnetic or electric dipoles result in spintronic and ferroelectric electronic devices. Similarly, the manipulation and detection of polar order in the ferroelectric-like metals might also generate promising electronic devices. However, the screen effect from the conduction electrons seems to prohibit the electrical switching and detection of the polar order in the ferroelectric-like metals, which limits the electronic applications of these materials. Theoretical proposals have been made for the electrical manipulation of the ferroelectric-like metals, such as decreasing the ferroelectric-like metal thickness \cite{RN2408}, using the interface coupling between the ferroelectric-like metal and the ferroelectric insulating substrate \cite{RN2407, RN2526}, \emph{etc.}. Experimentally, the switching of the ferroelectric-like metal has been demonstrated in few-layers WTe$_2$ by using a double gate device \cite{RN2306} and in bulk WTe$_2$ single crystal using a piezoresponse force microscopy \cite{RN2355}. Despite these remarkable progresses, the efficient detection of the polar order in ferroelectric-like metal remains a challenging problem.

Our calculation shows the nonlinear Hall effect can reflect not only the ferroelectric-like phase transition but also the polar order of LiOsO$_3$. If ${{d}_{z}}$ is positive for the positive polar displacement ($\delta>0$), a positive Hall voltage appears along the $z$-direction, as shown in Fig. \ref{FigIlldomain}(a). On the other hand, the switching of the polar direction is equivalent to apply the inversion symmetry operation in LiOsO$_3$, which changes the sign of the Berry curvature dipole and thus reverses the nonlinear Hall voltage [Fig. \ref{FigIlldomain}(c)]. However, the paraelectric-like state cannot generate the nonlinear Hall signal due to the symmetry. This strong dependence of the nonlinear Hall voltage on the polar direction is expected to be robust against the unavoidable doping effect by defects, because the sign of ${{d}_{z}}$ remains invariable in a broad energy window of 40 meV around $E_F$ (Fig. \ref{FigDxy}). Furthermore, as shown in Eq. (\ref{Eq_NHE}), the nonlinear Hall current introduced by a high-frequency electric field can be decomposed into a rectified current and a second harmonic one. It allows to distinguish the output signal from the input current by frequency easily. In addition to LiOsO$_3$, other ferroelectric-like metals such as Ca$_3$Ru$_2$O$_7$ \cite{RN2357}, Cd$_2$Re$_2$O$_3$ \cite{RN2386, RN2314}, GeTe \cite{RN2387}, NdNiO$_3$ \cite{RN2400} \emph{etc.} \cite{RN2299, RN2406, RN2409} are promising candidates to host nonlinear Hall effect. The manipulation and detection of polar order in the ferroelectric-like metal will be beneficial to expand the corresponding electronic applications.

\section{Conclusion}
\label{sec:Conclusion}
In conclusion, we propose that the nonlinear Hall effect can exist in ferroelectric-like metals. As a representative example, we consider the most investigated ferroelectric-like metal LiOsO$_3$ and showe that it has a large Berry curvature dipole, which can introduce a sizable nonlinear Hall effect. This effect disappears in the paraelectric-like phase of LiOsO$_3$. Moreover, the strong dependence of the Berry curvature dipole on the polar displacement offers an efficient method to detect the polar order. Therefore, the nonlinear Hall effect can be used in the study of the ferroelectric-like structural phase transition and expands potential applications in ferroelectric-like electronic devices. We hope our predictions will stimulate the experimental exploration of the nonlinear Hall effect in ferroelectric-like metals.

\begin{acknowledgments}
This work is supported by the NBRPC under No. 2019YFA0308403, National Nature Science Foundation of China under Grants No. 11947212, No. 11822407, No. 11534001, and Postdoctoral Science Foundation No. 2018M640513. We thank Prof. Bin Xu for helpful discussion.
\end{acknowledgments}

\end {document}